\documentclass[aps,pre,showpacs,amsmath,amsfonts,amssymb,superscriptaddress]{revtex4}

\usepackage{isolatin1}
\usepackage{psfig}
\usepackage[dvips]{graphicx}

\begin{document}

\title{Weak and strong chaos in FPU models and beyond}

\author{Marco Pettini}
\altaffiliation[also at: ] {Centro Interdipartimentale per lo
Studio delle Dinamiche Complesse (CSDC), Università di Firenze,
Istituto Nazionale per la Fisica della Materia (INFM), Unità di
Ricerca di Firenze, and Istituto Nazionale di Fisica Nucleare
(INFN), Sezione di Firenze, via G.~Sansone 1, I-50019 Sesto
Fiorentino (FI), Italy} \email{pettini@arcetri.astro.it}
\affiliation{INAF - Osservatorio Astrofisico di Arcetri, Largo
E.~Fermi 5, I-50125 Firenze, Italy}

\author{Lapo Casetti}
\altaffiliation[also at: ] {Centro Interdipartimentale per lo
Studio delle Dinamiche Complesse (CSDC), Universit\`a di Firenze,
Istituto Nazionale per la Fisica della Materia (INFM), Unit\`a di
Ricerca di Firenze, and Istituto Nazionale di Fisica Nucleare
(INFN), Sezione di Firenze, via G.~Sansone 1, I-50019 Sesto
Fiorentino (FI), Italy} \email{casetti@fi.infn.it}
\affiliation{Dipartimento di Fisica, Università di Firenze, via
G.~Sansone 1, I-50019 Sesto Fiorentino (FI), Italy}

\author{Monica Cerruti-Sola}
\email{mcs@arcetri.astro.it} \affiliation{INAF - Osservatorio
Astrofisico di Arcetri, Largo E.~Fermi 5, I-50125 Firenze, Italy}

\author{Roberto Franzosi}
\email{franzosi@df.unipi.it} \affiliation{Dipartimento di Fisica,
Università di Pisa, via Buonarroti 2, I-56127 Pisa, Italy}

\author{E.\ G.\ D.\ Cohen}
\email{egdc@rockefeller.edu} \affiliation{The Rockefeller
University, 1230 York Avenue, New York, NY 10021-6399, USA}

\date {\today}

\begin{abstract}
We briefly review some of the most relevant results that our group
obtained in the past, while investigating the dynamics of the
Fermi-Pasta-Ulam (FPU) models. A first result is the numerical
evidence of the existence of two different kinds of transitions in
the dynamics of the FPU models: {\it i)} a Stochasticity Threshold
(ST), characterized by a value of the energy per degree of freedom
below which the overwhelming majority of the phase space
trajectories are regular (vanishing Lyapunov exponents). It tends
to vanish as the number $N$ of degrees of freedom is increased.
{\it ii)} a Strong Stochasticity Threshold (SST), characterized by
a value of the energy per degree of freedom at which a crossover
appears between two different power laws of the energy dependence
of the largest Lyapunov exponent, which phenomenologically
corresponds to the transition between {\it weakly} and {\it
strongly} chaotic regimes. It is stable with $N$. A second result
is the development of a Riemannian geometric theory to explain the
origin of Hamiltonian chaos. The starting of this theory has been
motivated by the inadequacy of the approach based on homoclinic
intersections to explain the origin of chaos in systems of
arbitrarily large $N$, or arbitrarily far from
quasi-integrability, or displaying a transition between weak and
strong chaos.
 Finally, a third result stems from the search for the transition
 between weak and strong chaos in systems other than FPU.
 Actually, we found that a very sharp SST appears as the dynamical
counterpart of a thermodynamic phase transition, which in turn has
led, in the light of the Riemannian theory of chaos, to the
development of a topological theory of phase transitions.

\end{abstract}

\maketitle

\textbf{In a foreword to their co-authored work reprinted in the
Fermi Collected Papers \cite{FPU}, S.\ Ulam wrote: ``...Fermi
expressed often the belief that future fundamental theories in
physics may involve non-linear operators and equations, and that
it would be useful to attempt practice in the mathematics needed
for the understanding of nonlinear systems. The plan was then to
start with the possibly simplest such physical model and to study
the results of the calculation of its long-time behavior.... The
motivation then was to observe the rates of mixing and
thermalization with the hope that the computational results would
provide hints for a future theory. One could venture a guess that
one motive in the selection of problems could be traced to Fermi's
early interest in the ergodic theory...''}

\textbf{Actually, Fermi's early interest in ergodic theory is
witnessed by his contribution to a theorem due to Poincar\'e and
thenceforth known as the Poincar\'e-Fermi theorem. This asserts
that neither analytic (Poincar\'e) nor smooth (Fermi) integrals of
motion besides the energy can survive a generic perturbation of an
integrable system with three or more degrees of freedom, thus, in
the absence of other isolating integrals of motion, any constant
energy surface of these generic systems is expected to be
everywhere accessible to the phase space trajectory. At this
level, no hindrance to ergodicity seems to be possible. Whence the
surprise of Fermi, Pasta and Ulam (FPU) when no apparent tendency
to equipartition was observed in their numerical experiment whose
50th anniversary we are celebrating. Fermi himself considered what
they found a ``little discovery''. The almost contemporary
announcement by Kolmogorov of the starting of what would later 
become the celebrated KAM theorem, seemed to provide an
explanation to the unexpected FPU's results. But later
developments of KAM theory, including optimal estimates of the
$N$-dependence of the perturbation thresholds and the Nekhoroshev
theorem, revealed that this is not really an adequate framework to
explain the FPU problem. The rich variety of the numerical
phenomenology accumulated over time seemed to keep off "the
hope that the computational results would provide hints for a
future theory". In fact, "rates of mixing and thermalization" have
a startling and complicated dependence on energy, number of degrees
of freedom and initial conditions. Actually, any dynamical
evolution of the system depends on the starting point in phase
space and on the "landscape" of its surroundings. Thus, there can
be a huge variety of dynamical behavior entailed by the
preparation of the system in an initial condition out of
equilibrium. As a consequence, in order to get some global
information on the phase space structure, independently of the
initial conditions, one has to look at the chaotic component of
phase space. This way of tackling the FPU problem is very illuminating
and leads to the conclusion that the FPU problem does not threaten
the validity of statistical mechanics. Moreover, this has
stimulated the starting of a new theory of Hamiltonian chaos. }

\section{Introduction}

 Few problems have been studied so extensively over the last decades as the
one devised originally by E.\ Fermi, J.\ Pasta and S.\ Ulam (FPU)
in 1954 \cite{FPU}. Their purpose was to check numerically that a
generic but very simple non-linear many particle dynamical system
would indeed behave for large times as a statistical mechanical
system, that is it would approach equilibrium if initially not in
equilibrium. In particular their purpose was to obtain the usual
equipartition of energy over
 all the degrees of freedom of a system,
 for generic initial conditions.
To their surprise, for the system FPU considered -- a one
dimensional anharmonic chain of 32 or 64 particles with fixed ends
and in addition to harmonic, cubic ($\alpha$-model) or quartic
($\beta$-model) anharmonic forces between nearest neighbors --
this was not observed. A variety of manifestly non-equilibrium and
non-equipartition behaviors was seen, including quasiperiodic
recurrences to the initial state. In fact, a behavior reminiscent
of that of a dynamical system with few degrees of freedom was
found, rather than the expected statistical mechanical behavior.
The duration of their calculations varied between 10000 and 82500
computation steps.
 These results raised the fundamental question about the validity or at least
the generally assumed applicability of statistical mechanics to
non-linear systems of which the system considered by FPU seemed to
be a typical example. Most of the attempts to clarify this
difficulty have approached the problem as one in dynamical systems
theory. These analyses have revealed many very interesting
properties of the FPU system and uncovered a number of possible
explanations for the resolution of the observed conflict with
statistical
 mechanics. The classical explanations are: {\it i)} the survival of invariant
tori in the phase space of a quasi-integrable system (KAM theory)
\cite{KAM}, {\it ii)} the existence of Zabusky and Kruskal's
solitons in a special continuum limit leading to the integrable
Korteweg de Vries equation \cite{Zabusky,Zabusky1}, {\it iii)} the
existence of an order-to-chaos transition \cite{Izrailev}.

In this paper we will first try to exhibit the reasons why this
apparently bona fide statistical mechanical system did not behave
as such and, in particular, what in our opinion the significance
of this apparent failure is for the general validity of
statistical mechanics; then, we will review a number of results
that we obtained studying the dynamics of the FPU models or being
directly inspired by it.

There are a number of obvious questions related to the
unstatistical mechanical behavior observed by FPU, which all
address the generic nature of the results of Fermi and
collaborators:

\begin{description}
\item{a)} Was their time of integration long enough?
\item{b)} Was their dynamical system of $N=32$ or $64$ particles in one
dimension large enough, i.e. possessing a sufficient number of
degrees of freedom, to qualify as a statistical mechanical system?
\item{c)} Were the recurrence phenomena (to within
$3
\% 
$) observed by FPU, transient or generic, i.e., possibly
related to a Poincar\'e recurrence time?
\end{description}

The search for answers to these questions made the work of FPU
very seminal, spawning many new developments and connections in
the theory of nonlinear dynamical systems, such as the connection
with continuum models based on the Kor\-teweg -- de Vries
equation, leading to solitons \cite{Zabusky}, heavy breathers
etc., or with few degrees of freedom models like the
H\'enon-Heiles \cite{Henon} and the Toda lattice \cite{Lunsford}.

Thus, the effort to resolve the so-called FPU problem has led to
enormous advances in our understanding of non linear dynamical
systems; for a review we refer to \cite{Ford}. Although in our
opinion the FPU problem has possibly not yet exhausted its power
of inspiration, we believe that the FPU paradox, i.e., FPU's
original question, can nowadays be reasonably answered along the
lines we are going to describe hereafter.

\section{Stochasticity thresholds in FPU models}

For some time, it has been a commonly accepted idea \cite{Ford}
that the KAM theorem provides
 the essential answer to FPU's observations, i.e., for sufficiently small
nonlinearities and a class of initial conditions living on
non-resonant tori, the FPU system behaves like
 an integrable system and is represented by deformed tori in phase space. With
increasing strength of the non-linearities, a progressive chaotic
behavior appears, which would ultimately lead to the expected
approach to equilibrium and equipartition.
Even though we found regular regions in phase space, the existing
typical KAM estimates of the $N$-dependence of the perturbation
threshold (below which a positive measure of KAM tori survive),
are qualitatively different from our results, indicating that the
physics of the FPU model is quite different from what is contained
in these estimates.

Thanks to the power of modern computers, we have considerably
extended the calculations performed in the past by various authors
and we have been able to reconcile different, and sometimes
contradictory, aspects of the FPU dynamics.

\subsection{FPU-$\alpha$ model}
Very interesting results have been obtained revisiting the
FPU-$\alpha$ model \cite{noi} by focusing on the development of
chaoticity in the time evolution of the system rather than on the
attainment of equipartition. For these numerical experiments, the
chosen initial conditions -- single mode excitations -- were the
same as chosen by Fermi and collaborators in their original
experiment.

The model is described by the Hamiltonian \cite{FPU}
\begin{equation}
H({ p},{ q}) = \sum_{k=1}^N\left[ \frac{1}{2}p_k^2 + \frac{1}{2}
(q_{k+1} - q_k)^2 + \frac{\alpha}{3}(q_{k+1} - q_k)^3\right],
\label{HFPUalpha}
\end{equation}
where the particles have unit mass and unit harmonic coupling
constant and the end-points are fixed ($q_1=q_{N+1}=0$).

Comparing the behaviour in time of the largest Lyapunov exponent
in the FPU system with that of the same quantity in a suitable
integrable system, it has been possible to define clearly what a
trapping time in a regular region of phase space is and to
determine numerically and unambiguously its value. The integrable
system we chose for this comparison is the Toda lattice, from
which the FPU-$\alpha$ model can be obtained as a third order
truncation of the power series expansion of its potential. The
Toda lattice is defined by the Hamiltonian
\begin{equation}
H(p, q)=\sum_{k=1}^N\frac{1}{2}p_k^2 + \frac{a}{b}\sum_{k=1}^N
\left[ e^{-b(q_{k+1}-q_k)}+b(q_{k+1}-q_k)-1\right]~. \label{Htoda}
\end{equation}
The decay pattern toward zero of $\lambda^{Toda}(t)$ is
undistinguishable from the decay pattern of $\lambda^{FPU}(t)$ up
to some time $\tau_T$, after which $\lambda^{FPU}(t)$ separates
from $\lambda^{Toda}(t)$ and converges to a finite value while
$\lambda^{Toda}(t)$ goes to zero. This suggests that
non-integrable motions of the FPU lattice, originated by one-mode
initial excitations, enter their chaotic component after a
transient and possibly long trapping in a regular region of phase
space (some kind of ``Nekhoroshev-like'' trapping) and that
equipartition is eventually attained on a finite, albeit possibly
very long, time scale.

\begin{figure}[ht]
\centerline{\psfig{file=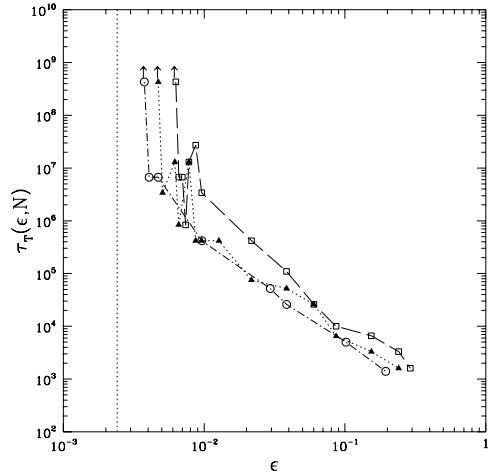,height=8cm,clip=true}}
\caption{ FPU-$\alpha$ model. The trapping times
$\tau_T(\varepsilon,N)$ at different values of energy density
$\varepsilon$ (i.e. at different values of the initial excitation
amplitudes), are reported. Open squares refer to the case $N=32$,
solid triangles refer to $N=64$, open circles refer to $N=128$,
respectively. The endpoints of the broken lines are lower bounds
for the trapping time (the cut-off of the integration time is at
$t=4.3 \times 10^{8}$). The dotted vertical line at
$\varepsilon=0.00241$ corresponds to the initial excitation
amplitude of the FPU's original paper. From Ref.
\protect\cite{noi}.\label{fig_cohen_3}}
\end{figure}

Fig.\ \ref{fig_cohen_3} shows that the trapping times
$\tau_T(\varepsilon ,N)$ -- so defined -- for the FPU-$\alpha$
model at different values of both the energy density $\varepsilon$
and of the number of degrees of freedom $N$, with decreasing
$\varepsilon$ first tend to increase monotonically, then,
abruptly, display an apparently divergent behavior.


This very steep increase of $\tau_T$ with decreasing $\varepsilon$
suggests the existence of, at least, a very narrow bottleneck in
phase space, through which the system can only escape with great
difficulty or, perhaps, it might not escape at all. The sharp
variation with $\varepsilon$ of the shape of $\tau_T(\varepsilon)$
brings about a natural definition of a threshold value of
$\varepsilon$ below which $\tau_T$ seems to diverge.

\begin{figure}[ht]
\centerline{\psfig{file=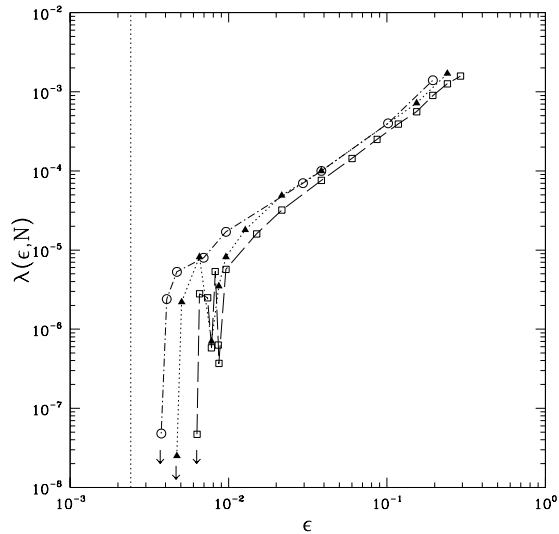,height=8cm,clip=true}}
\caption{ FPU-$\alpha$ model. The largest Lyapunov exponents
$\lambda(\varepsilon,N)$ are shown for
 different values of the energy density $\varepsilon$ and a sine wave
initially excited. Symbols as in Fig.~\protect\ref{fig_cohen_3};
here the arrows are upper bounds for $\lambda$. From Ref.
\protect\cite{noi}. \label{fig_cohen_4}}
\end{figure}

For what concerns the behavior of the largest Lyapunov exponents,
when $\varepsilon$ is smaller than the threshold value,
$\lambda^{Toda}(t)$ and $\lambda^{FPU}(t)$ do not show any
separation, even after a very long integration time. That $\tau_T$
is really a trapping time and not, for example, a trivial effect
of the numerical statistics is suggested by the fact that
$\tau_T(\varepsilon)\sim\varepsilon^{-2}$ whereas
$\lambda(\varepsilon)\sim\varepsilon^{3/2}$, that is
$\lambda\neq\tau_T^{-1}$. In Fig.\ \ref{fig_cohen_4},
$\lambda^{FPU}(\varepsilon ,N)$ is reported.


The shapes of both $\lambda^{FPU}(\varepsilon ,N)$ and
$\tau_T(\varepsilon ,N)$ strongly suggest the existence of a
threshold value -- {\it which depends on} $N$ -- of the energy
density, above which the motion is chaotic and below which the
trajectories appear to belong to a regular region of phase space.
This threshold is referred to as stochasticity threshold (ST). To
our knowledge, its {\it direct} evidence in a non-linear
Hamiltonian system at $N\gg 2$ has been found for the first time
in \cite{noi}.

Fermi and coworkers chose an initial condition well below this ST
(the energy density corresponding to their initial condition is
shown by the vertical dotted line in Fig.\ \ref{fig_cohen_4}); had
they taken a ten times larger amplitude of the initial excitation,
they would have observed equipartition during the integration time
they used. This appears to be the simple but non-trivial
explanation of the lack of statistical mechanical behavior
observed in the original FPU numerical experiment.


In order to understand whether the ST refers to a global property
of the constant energy surface $\Sigma_E$ or is rather a local
property of $\Sigma_E$, sensitive to the initial condition, two
other choices of more physically generic initial conditions, i.e.,
random positions and momenta, were considered.

In each case a threshold energy (or equivalently energy density,
since $N$ is fixed) was found. This fact suggests that the phase
space undergoes some important structural change as the energy is
varied: we can find regions of the phase space where ordered
trajectories are observed, regions where there is coexistence of
order and chaos and regions where chaos is fully developed.

\begin{figure}[ht]
\centerline{\psfig{file=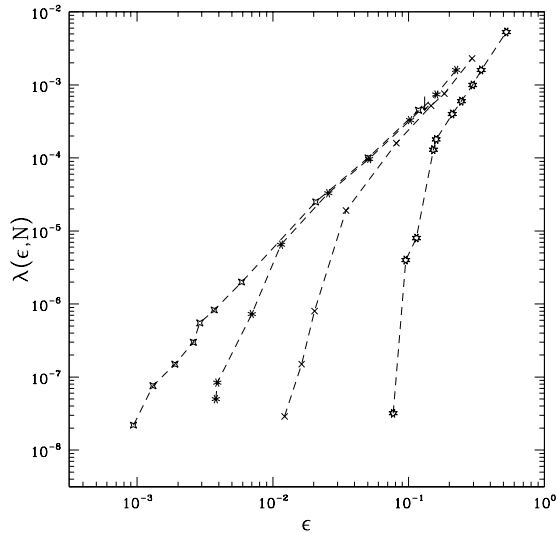,height=8cm,clip=true}}
\caption{ FPU-$\alpha$ model. The largest Lyapunov exponents
$\lambda (\varepsilon,N)$ are plotted {\it vs.} the energy density
$\varepsilon$, for different values of $N$. Random initial
conditions are chosen.
 Star-like polygons refer to $N=8$, crosses to $N=16$, asterisks to $N=32$
 and star-like squares to $N=64$, respectively.
The arrows have the same meaning as in Fig.
\protect\ref{fig_cohen_4}. From Ref. \protect\cite{noi}.
\label{fig_cohen_6}}
\end{figure}

An important question is whether the stochasticity threshold is
stable or unstable with respect to $N$. Unambiguous information
about this point is provided by the Lyapunov exponents
$\lambda(\varepsilon ,N)$ computed at different $N$, always
starting with random initial conditions (Fig.\ \ref{fig_cohen_6}).


At large $\varepsilon$, there is a tendency of all the sets of
points to join, while they tend to separate at small
$\varepsilon$: the larger $N$, the smaller the energy density at
which the separation occurs. The ``critical'' energy density
$\varepsilon_c$ at which the separation occurs shows the
$N$-dependence $\varepsilon_c(N)\propto 1/N^2$.

A qualitative agreement about the vanishing with $N$ of the
critical energy to get chaos is reported in a recent paper on the
FPU-$\alpha$ model \cite{Shepelyansky}. The question of how to
explain the existence and the $1/N^2$ dependence of the
stochasticity threshold remains open.

We thus see that revisiting the FPU-$\alpha$ model led to the
observation of some very interesting phenomena: the apparent
existence of regular regions in the phase space of a
non-integrable Hamiltonian system with many degrees of freedom at
large values of the anharmonic energy (even very large if compared
with what they should be according to the KAM theory), and the
existence of almost regular regions of phase space where the
trajectories are trapped during long but finite times. The
behavior of the largest Lyapunov exponent suggests that the sudden
escape from the regular region might occur as if the trajectory
would eventually find a 'hole' in its boundary.

Moreover, the coexistence of regular regions of the phase space
and of a large chaotic ``sea'' reconciles different and sometimes
apparently contradictory aspects of the FPU dynamics found in the
past. The lack of equipartition in the original FPU experiment is
not representative of a global property of phase space: apparently
regular, soliton-like structures, similar to those of Zabusky and
Kruskal, have a very long, possibly infinite, life-time below the
stochasticity threshold, whereas, above the same threshold, they
have only a finite life-time\cite{noi}.

The threshold energy density for the onset of chaos shows a clear
tendency to vanish at an increasing number of degrees of freedom
($\sim 1/N^2$), so that strong evidence has been found in support
of the point of view that the so-called ``FPU-problem'' does not
invalidate the (generic) approach to equilibrium and the validity
of equilibrium statistical mechanics. On the other hand the
existence of long living initial states far from equilibrium,
 may well have
interesting, non trivial physical applications.

\subsection{FPU-$\beta$ model}

The FPU-$\beta$ model is described by the Hamiltonian \cite{FPU}
\begin{equation}
H({ p},{ q}) = \sum_{k=1}^N\left[ \frac{1}{2}p_k^2 + \frac{1}{2}
(q_{k+1} - q_k)^2 + \frac{\beta}{4}(q_{k+1} - q_k)^4\right],
\label{HFPUbeta}
\end{equation}
where the particles have unit mass and unit harmonic coupling
constant and the end-points are fixed ($q_1=q_{N+1}=0$); for this
model also periodic boundary conditions have been considered
($q_1=q_{N+1}$).

The approach to equilibrium of the FPU-$\beta$ model was studied
extensively for various classes of initial conditions by Kantz
{\it et al.}\ \cite{KLR} and recently by De Luca {\it et al.}\
\cite{DLR} who extended and improved earlier computations of ours
\cite{sparpa1,sparpa2}.

A very detailed picture has emerged from these works, as to the
behavior of the FPU-$\beta$ model in its dependence on
non-equilibrium initial conditions as well as in the role played
by low frequency and
 high frequency mode-mode couplings
\cite{Poggi} during its time evolution.

\begin{figure}[ht]
\centerline{\psfig{file=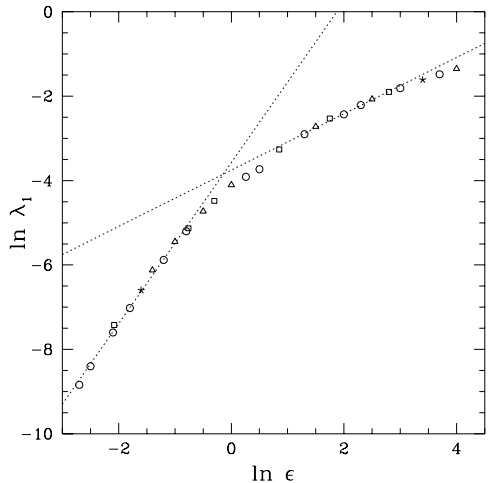,height=8cm,clip=true}}
\caption{FPU-$\beta$ model. Largest Lyapunov exponents
$\lambda_{1}$ vs. energy density $\varepsilon$ at $N=128$ and at
different initial conditions: random at equipartition (circles),
wave packets at different average wave numbers (squares, triangles
and asterisks). From Ref.~\protect\cite{PettiniCerruti}.
\label{fig_sst}}
\end{figure}

Several years ago, we introduced \cite{sparpa2} a time dependent
spectral entropy $S(t)=-\sum_i w_i(t) \log w_i(t)$, where
$w_i(t)=E_i(t)/\sum_k E_k(t)$ is the normalized energy content of
the $i$-th harmonic normal mode, defined so as to detect energy
equipartition (when it attains its maximum value) and to measure
the time needed to reach it. By means of this spectral entropy, we
investigated in Refs.\ \cite{PettiniLandolfi,PettiniCerruti} the
relationship between equipartition times, measured through the
time relaxation patterns of this spectral entropy, and the chaotic
properties of the dynamics in nonlinear large Hamiltonian systems.
For the FPU-$\beta$ model, we have put in evidence that, at
different initial conditions and at long times, the spectral
entropy always relaxes toward its maximum value signaling
equipartition, however, depending on the value of the total energy
density, the relaxation occurs with quite different modalities.
 The relaxation time is approximately constant for energy
densities greater than some threshold value ${\varepsilon}_c$, but
it steeply grows by decreasing the energy density below this
threshold. Moreover, the largest Lyapunov exponent shows a
crossover in its ${\varepsilon}$-dependence corresponding to this
threshold value. We interpret this phenomenological result as the
(smooth) transition -- at ${\varepsilon}_c$ -- between two
different regimes of chaoticity, weak chaos and strong chaos, thus
we called this transition the Strong Stochasticity Threshold (SST)
\cite{PettiniCerruti}. Weak and strong chaos are qualitative terms
to denote slow and fast phase space mixing respectively. In Refs.\
\cite{PettiniLandolfi,PettiniCerruti} we resorted to a random
matrix model for the tangent dynamics to try to make more precise
and quantitative the definitions of weak and strong chaos. At
least in a limited high energy range of values, the random matrix
model predicts the numerically observed scaling $\lambda
(\varepsilon)\sim\varepsilon^{2/3}$ (this law changes to $\lambda
(\varepsilon)\sim\varepsilon^{1/4}$ at very high energy density,
however this is not explained by the random matrix model, the
reason is that a free parameter, a time-scale of unknown
$\varepsilon$ dependence, enters the random matrix model. This
time-scale is arbitrarily assumed constant). Thus we say that {\it
chaos is strong} in the energy density range where $\lambda
(\varepsilon)\sim\varepsilon^{2/3}$, because the random matrix
model assumes that the dynamics looks as a random uncorrelated
process (if sampled with the just mentioned unknown time scale).
At low energy density, the $\varepsilon$-scaling of $\lambda$ is
found to be steeper, $\lambda (\varepsilon)\sim\varepsilon^{2}$,
so that $\lambda$ fastly decreases as ${\varepsilon}$ is lowered
and is much smaller than it should be if the high energy random
matrix prediction could be extrapolated down to low energy values.
For this reason we say that here {\it chaos is weak}. Figure
\ref{fig_sst} shows $\lambda (\varepsilon, N)$.


The SST is independent of the initial conditions so it has to be
ascribed to some change in the global properties of the phase
space, for this reason it has to have major consequences on the
dynamics. An interesting explanation based on a model for phase
space diffusion is given in Ref.\ \cite{Tsaur}.

The SST has been found to be correlated with changes in the {\it
transient non-equilibrium} behavior (e.g., relaxation to
equipartition) \cite{PettiniCerruti,PettiniLandolfi,Xenon}, and
has been found to be also correlated with {\it stationary
non-equilibrium} phenomena like heat
conduction\cite{livilepripoliti}. The SST is found to be
independent of the number of degrees of freedom, which makes it of
prospective relevance for {\it equilibrium statistical mechanics}.
Among the model dependent consequences of the existence of the
SST, it is worth mentioning that in the FPU-$\beta$ model, at
$\varepsilon <\varepsilon^{SST}_c$ high-frequency excitations
yield longer relaxation times with respect to low frequencies.
This is in agreement with the common belief that high-frequencies
have the tendency to freeze; at $\varepsilon >\varepsilon^{SST}_c$
the situation is reversed. High frequency excitations yield a
quicker relaxation than low frequencies \cite{PettiniCerruti}.

It is remarkable that the existence of the SST is not only a
characteristic of the FPU-${\beta}$ model. In fact, it has been
detected in the following one dimensional lattices: with diatomic
Toda interactions (i.e., with alternating masses that break
integrability) \cite{Yoshimura}; with single-well ${\phi^4}$
interactions \cite{PettiniLandolfi}; with smoothed Coulomb
interactions \cite{Yoshimura}; with Lennard-Jones interactions
\cite{Yoshimura}; in an isotropic Heisenberg spin chain
\cite{greci}; in a coupled rotators chain which displays two
thresholds separating two regions of weak chaos (occurring at low
and high energies) from an intermediate region of strong chaos
\cite{eklr,CCP}; in a ``mean-field'' XY chain \cite{Firpo} and in
homopolymeric chains \cite{omopolimeri}. It has been also detected
in two and three dimensional lattices, with two-wells $\varphi^4$
interactions \cite{CCP1,CCCPPG}, with XY Heisenberg interactions
\cite{Butera,CCCP}. Therefore the SST seems to be a generic
property of Hamiltonian systems with many degrees of freedom.

\subsection{FPU-($\alpha+\beta$) model}

\begin{figure}[ht]
\centerline{\psfig{file=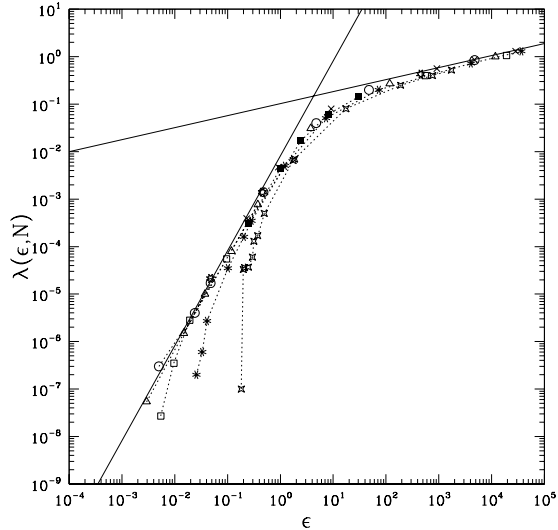,height=8cm,clip=true}}
\caption{FPU-($\alpha +\beta$) model.
 The largest Lyapunov exponents $\lambda(\varepsilon,N)$ are shown for
 different values of the energy density $\varepsilon$ for various values
of $N$. Starlike squares refer to $N=8$, asterisks to $N=16$,
 open squares to $N=32$, open triangles to $N=64$,
open circles to $N=128$, starlike polygons to $N=512$ and crosses
to $N=1024$, respectively. Full squares refer to $N=32$ and
excitation amplitudes $A$ ranging from 5 to 50. Solid lines are
the asymptotic scalings $\varepsilon^2$ and
$\varepsilon^{\frac{1}{4}}$ at low and high energy density,
respectively. From Ref. \protect\cite{CePeCo}. 
\label{Fig1}}
\end{figure}

In the FPU-$\alpha$ model, the existence of a stochasticity
threshold (ST) at an energy density below which the dynamics is
regular has been observed. In the FPU-$\beta$ model, a strong
stochasticity threshold (SST) above which the dynamics is strongly
chaotic has been found. By combining these two models into the
FPU-($\alpha + \beta$) model, it is possible to observe the
coexistence of both the ST and the SST. This model has been
studied recently in \cite{CePeCo}. It is described by the
Hamiltonian
\begin{eqnarray}
H({ p},{ q})& =& \sum_{k=1}^N\left[ \frac{1}{2}p_k^2 + \frac{1}{2}
(q_{k+1} - q_k)^2 + \frac{\alpha}{3}(q_{k+1} - q_k)^3
+ \frac{\beta}{4} (q_{k+1} - q_k)^4\right]~, \label{HFPU}
\end{eqnarray}
where the particles have unit mass and a unit harmonic coupling
constant and the end-points are fixed ($q_1=q_{N+1}=0$). This
model Hamiltonian, with the choice of $\alpha =0.25$ and $\beta
=\frac{2}{3} \alpha^2$, is a fourth-order expansion of the Toda
model (\ref{Htoda}). Consequently, its potential function is very
close to interatomic potentials of the Morse or Lennard-Jones type
in solids, provided that a suitably restricted energy density
range is considered. Random initial conditions have been chosen.
The results of the computation of the largest Lyapunov exponents
at different energy densities and for different values of $N$ are
shown in Fig. \ref{Fig1}. The patterns of $\lambda(\varepsilon
,N)$, therein reported, display some remarkable features. For
small values of the energy density, there is a sudden drop of
$\lambda$ which, in close analogy with Ref.\ \cite{noi}, allows us
to define an ST below which we can assume that the overwhelming
majority of the trajectories in phase space are regular. This ST
moves to smaller and smaller values of $\varepsilon$ as $N$ is
increased.


Around $\varepsilon \simeq0.8$, a ``knee'' is observed in the
pattern $\lambda(\varepsilon,N)$ (Fig.\ \ref{Fig1}), due to a
crossover between two power law behaviors, $\sim \varepsilon^2$ at
small $\varepsilon$ and
 $\sim \varepsilon^{\frac{1}{4}}$ at
large $\varepsilon$, where the latter has been attributed to the
existence of an SST \cite{PettiniLandolfi,PettiniCerruti}. This
crossover is the signature of the transition from weak to strong
chaos, as already discussed in
\cite{PettiniLandolfi,PettiniCerruti}.

\section{Riemannian geometry of chaos in the FPU-$\beta$ model}
\label{geometria}

In this Section we sketch how we have analytically computed, in
the limit of arbitrarily large $N$, the largest Lyapunov exponent
$\lambda$ as a function of the energy density $\varepsilon$ for
the FPU-$\beta$ model. The excellent agreement of the analytic
outcome with the numerical results for $\lambda(\varepsilon)$
provides a preliminary understanding of the transition between
weak and strong chaos (SST), and strongly supports the general
validity of the proposed explanation of the origin of Hamiltonian
chaos.

For generic non-integrable Hamiltonian systems, when the number of
degrees of freedom is large, which in practice means already a few
hundred, the whole phase space is filled by chaotic trajectories,
at least at physically meaningful values of the energy density.
Therefore, any framework of analytic description of the dynamics
has to cope with chaos. However, even the basic question about the
origin of chaos itself, in many degrees of freedom Hamiltonian
systems seems to lack an answer. For example, all the theoretical
machinery of Classical Perturbation Theory (CPT) is of little use
if we want to deal with chaos, and so does the traditional
explanation of its origin based on {\it homoclinic intersections}
\cite{CCP}.

Until a few years ago, the ``only game in town'', which seemed of
potential interest to treat chaos at large $N$, was an attempt by
Krylov \cite{Krylov} at explaining the origin of phase space
mixing as a consequence of negative scalar curvature of suitable
Riemannian manifolds whose geodesics coincide with the solutions
of Newton equations of motion. Krylov's idea was to take advantage
of some mathematical results about the stability properties of
geodesics on negatively curved Riemannian manifolds. These results
are associated with the names of Hadamard \cite{Hadamard}, Hedlund
\cite{Hedlund} and Hopf \cite{Hopf}. Since Krylov's, other
attempts have been done along the same line of thought (see e.g.
the discussion in \cite{physrep}), but none of them appeared very
useful.

More recently, we have reconsidered the Riemannian geometric
approach and, with the aid of numerical simulations on the FPU-$\beta$
model, we have discovered why the previous attempts
failed: the dominant mechanism for chaotic instability in
physically relevant geodesics flows is {\it parametric
instability} due to curvature variations along the geodesics, and
 {\it not necessarily geodesic flows on negatively curved manifolds}
\cite{Pettini,CerrutiPettini1,CerrutiPettini2,PettiniValdettaro,CCP,physrep}.
On this basis, we have started the formulation of a Riemannian
theory of Hamiltonian chaos which applies to dynamical systems
described by a standard Lagrangian function
\begin{equation}
L({ q},{\dot q})=\frac{1}{2}a_{ik}\dot q^i\dot q^k -V({ q})\ ,
\label{lagr}
\end{equation}
where $a_{ik}$ is the kinetic energy matrix ($a_{ik}=\delta_{ik}$
for the usual form of the kinetic energy) or, equivalently, by the
Hamiltonian
$H({ p}, { q})=\frac{1}{2}a^{ik}p_ip_k + V({ q})$,
where the momenta are given by $p_i=a_{ik}\dot q^k$.
 From Maupertuis' least action principle for asynchronous isoenergetic varied
 paths $\gamma(t)$ with fixed endpoints
\begin{equation}
\delta\int_{\gamma}2W({ q}, {\dot q}) d\, t =
\delta\int_{\gamma}\{2 [E - V({ q})]a_{ik}\dot q^i\dot
q^k\}^{1/2}dt = 0\ , \label{leastaction}
\end{equation}
where $W$ is the kinetic energy, the equations of motion follow.
Equation (\ref{leastaction}) is equivalent to the extremization of
the length-integral $\int_\gamma ds$ where $ds$ is $ds^2=g_{ik}({
q})dq^idq^k=2[E-V({ q})]a_{ik}dq^idq^k$. In other words,
mechanical trajectories are geodesics of the configuration space
endowed with a proper Riemannian manifold structure described by
the metric tensor
\begin{equation}
g_{ik}({ q})=2[E-V({ q})]a_{ik}~. \label{Jmetric}
\end{equation}
This is known as Jacobi metric and is defined in the region of the
configuration space where $V({ q})$. In local coordinates, the
geodesic equations on a Riemannian manifold are given by
\begin{equation}
{\frac{d^2 q^i}{ds^2}}+
\Gamma^i_{jk}{\frac{dq^j}{ds}}{\frac{dq^k}{ds}}=0\ , \label{geo}
\end{equation}
where $s$ is the proper time and $\Gamma^i_{jk}$ are the
Christoffel coefficients of the Levi-Civita connection associated
with $g_{ik}$. By direct computation, using $g_{ik}=(E-V({
q}))\delta_{ik}$, $\Gamma^i_{jk}=
{\frac{1}{2W}}\delta^{im}(\partial_jW\delta_{km}+\partial_kW
\delta_{mj}-\partial_mW\delta_{jk})$ and $ds^2=2W^2dt^2$, it can
be easily verified that the geodesic equations yield
\begin{equation}
{\frac{d^2q^i}{dt^2}}= -{\frac{\partial V}{\partial
q^i}}\ ,\label{newton}
\end{equation}
i.e. Newton's equations associated to the Lagrangian
(\ref{lagr}).

Among other Riemannian geometrizations of Newtonian dynamics, a
very interesting one is defined in an enlarged configuration
spacetime $M\times {\mathbb{R}}^2$, with local coordinates
$(q^0,q^1,\ldots,q^i,\ldots,q^N,q^{N+1})$, endowed with a
non-degenerate pseudo-Riemannian metric whose arc-length is
\cite{Eisenhart}
\begin{equation}
ds^2 = g_{\mu\nu}\, dq^{\mu}dq^{\nu} = a_{ij} \, dq^i dq^j -2V({
q})(dq^0)^2 + 2\, dq^0 dq^{N+1} ~, \label{g_E}
\end{equation}
called {\em Eisenhart metric}. The natural motions are obtained as
the canonical projection on the configuration space-time of those
geodesics for which the arclength is positive-definite and given
by
$ds^2 = ({\rm const} )^2 dt^2$.
A way of measuring of how much a Riemannian manifold deviates from
being a Euclidean manifold is provided by the degree of
non-commutativity of the covariant derivatives which is properly
defined by the Riemann-Christoffel curvature tensor
$R(X,Y)=\nabla_X\nabla_Y-\nabla_Y\nabla_X$,
where $\nabla$ is the Levi-Civita connection, and $X,Y$ are
tangent vectors \cite{DoCarmo}. There are two relevant curvature
scalars: the {\em Ricci curvature} $K_R$ in a given direction $v$,
and the {\em scalar curvature} ${\cal R}$ (see \cite{physrep}).

There is an important relation between the curvature of a manifold
and the stability of its geodesics. In fact, the evolution of a
vector field $J$, called geodesic separation vector, is completely
determined by the curvature tensor according to the equation
\begin{equation}
\nabla^2_{\dot{\gamma}}J(s)\, +\,
R[J(s),\dot{\gamma}(s)]\dot{\gamma}(s)\, =\,0.\label{eq_jacobi}
\end{equation}
This is the {\em Jacobi -- Levi-Civita equation}, where
$\nabla_{\dot{\gamma}}$ is the covariant derivative in the
direction of the velocity vector $\dot{\gamma}=v$. $J$ contains
the whole information on the stability -- or instability -- of any
given reference geodesic $\gamma (s)$ because it locally measures
the distance from $\gamma (s)$ of any given geodesic close to
$\gamma (s)$.

Since the Jacobi equation (\ref{eq_jacobi}) relates the stability
of the geodesics of a manifold to its curvature, the Jacobi
equation links stability and instability (chaos) of the dynamics
with the curvature of the ``mechanical'' manifold \cite{CFP},
 if the metric is associated with a physical system.

In the particular case of {\em isotropic} -- or {\em constant
curvature} -- manifolds, Eq.\ (\ref{eq_jacobi}) becomes very
simple: choosing a geodesic frame, i.e., a reference frame
transported parallel along a reference geodesic, the Jacobi
equation is written as
\begin{equation}
\frac{d^2 J}{ds^2} + K \, J = 0~, \label{eq_jacobi_cost}
\end{equation}
where $K=K_R/(N-1)\equiv {\cal R}/N(N-1)$, which has either
bounded oscillating solutions $\Vert J\Vert \propto\cos(\sqrt{K}\,
s)$ or exponentially unstable solutions $\Vert J\Vert \propto
\exp(\sqrt{-K}\, s)$ according to the sign of the curvature and
thus of the constant $K$.

As long as the curvatures are negative, the geodesic flow is
unstable even if the manifold is no longer isotropic, and the
instability exponent is greater than or equal to $(-
\max_M(K))^{1/2}$. Geodesic flows on compact manifolds with
everywhere negative curvature were studied for the first time in
the classic works by Hadamard, Hedlund and Hopf
\cite{Hadamard,Hedlund,Hopf} and many results were then
established by Anosov \cite{Anosov}, among them the fact that such
systems are ergodic and mixing.

Equation (\ref{eq_jacobi_cost}) is valid only when $K$ is
constant. Nevertheless in the case in which $\dim M = 2$
(surfaces), the Jacobi equation -- again written in a geodesic
reference frame for the sake of simplicity -- takes a form very
close to that of isotropic manifolds,
\begin{equation}
\frac{d^2 J}{ds^2} + \frac{1}{2} {\cal R}(s)\, J = 0~,
\label{eq_jacobi_2}
\end{equation}
where ${\cal R}(s)$ denotes the scalar curvature of the manifold
along the geodesic $\gamma(s)$. The solutions of Eq.
(\ref{eq_jacobi_2}) may exhibit an exponentially growing envelope
even if the curvature ${\cal R}(s)$ is everywhere positive but non
constant. For example, in the case of the celebrated
H\'enon-Heiles model \cite{Henon}, the scalar curvature ${\cal
R}$, computed with the Jacobi metric, is always positive despite
the existence of fully developed chaos above some threshold energy
\cite{CerrutiPettini2}. As a matter of fact, the generic condition
of physically relevant systems (like coupled anharmonic
oscillators on $d$-dimensional lattices) is that Ricci and scalar
curvatures of the mechanical manifolds are neither constant nor
everywhere negative, and the straightforward approach based on Eq.
(\ref{eq_jacobi_cost}) does not apply.

The key point is to realize that negative curvatures are not
necessary to generate chaos, while the generic non constancy of
the curvature of mechanical manifolds (in the absence of very
"exotic" hidden symmetries \cite{killing}), triggers {\em
parametric instability} of the geodesics. Thus the exponential
growth of the solutions of the stability equation
(\ref{eq_jacobi_2}), that is chaos, even if no negative curvature
is ``felt'' by the geodesics.

\subsection{A geometric formula for the Lyapunov exponent}

In the large $N$ case, with some simplifying assumptions, mainly
that the mechanical manifolds are {\it quasi-isotropic}
\cite{CCP,physrep}, it is possible to derive an effective scalar
stability equation resembling Eq.\ (\ref{eq_jacobi_2}), where the
role of ${\cal R}(s)$ is played by a random process, so that an
analytic estimate of the largest Lyapunov exponent can be worked
out.
 This effective equation is {\em independent of the knowledge of the dynamics}
and has the form
\begin{equation}
{\frac{d^2\psi}{ds^2}}+ \Omega (s)\, \psi =0 \label{eq_stoc_osc}
\end{equation}
where $\psi$ denotes any of the components of $J$ because now all
of them obey the same effective equation of motion, and the
squared frequency $\Omega (s)$ is a gaussian random process
\begin{equation}
\Omega (s) = \langle k_R\rangle_\mu\ + \langle\delta^2
k_R\rangle_\mu^{1/2}\,\eta (s)\ ,
\end{equation}
where $k_R=K_R/(N-1)$ and $\langle \delta^2 k_R \rangle_\mu$ is a
shorthand for $\frac{1}{N-1} \langle \delta^2 K_R \rangle_\mu$;
the averages $\langle \cdot \rangle_\mu$ are microcanonical
averages; $\eta (s)$ is a gaussian random process with zero mean
and unit variance.
 Our estimate for the (largest) Lyapunov exponent $\lambda$
is then given by the growth-rate of $\Vert
(\psi,\dot\psi)(t)\Vert^2 $ according to the definition
\begin{equation}
\lambda = \lim_{t\to\infty} \frac{1}{2t} \log \frac{\psi^2(t) +
\dot\psi^2(t)}{\psi^2(0) + \dot\psi^2(0)}~.
\label{def_lambda_gauss}
\end{equation}

The ratio $(\psi^2(t) + \dot\psi^2(t))/(\psi^2(0) +
\dot\psi^2(0))$ is computed by means of a technique developed by
Van Kampen, summarized in Ref.\ \cite{CCP}, where the following
expression for $\lambda$ has been derived

\begin{eqnarray}
\lambda(\Omega_0,\sigma_\Omega,\tau) & = & \frac{1}{2}
\left(\Lambda-\frac{4\Omega_0}
{3 \Lambda}\right), \nonumber \\
\Lambda & = &
\left(2\sigma^2_\Omega\tau+\sqrt{\left(\frac{4\Omega_0}{3}
\right)^3+(2\sigma^2_\Omega\tau)^2}\,\right)^{1/3},
\label{lambda_gauss}
\end{eqnarray}
where $\Omega_0=\langle k_R\rangle_\mu$, $\sigma_\Omega
=\langle\delta^2k_R\rangle_\mu =\frac{1}{N}[\langle
K_R^2\rangle_\mu -\langle K_R\rangle_\mu^2]$ and $\tau$ is a time
scale expressed in terms of $\Omega_0$ and $\sigma_\Omega$. The
quantities $\Omega_0$, $\sigma_\Omega$ and $\tau$ can be computed
as static, i.e. {\it microcanonical} averages. Therefore Eq.
(\ref{lambda_gauss}) gives an analytic, though approximate,
formula for the largest Lyapunov exponent {\it independently} of
the numerical integration of the dynamics and of the tangent
dynamics.

\begin{figure}[ht]
\centerline{\psfig{file=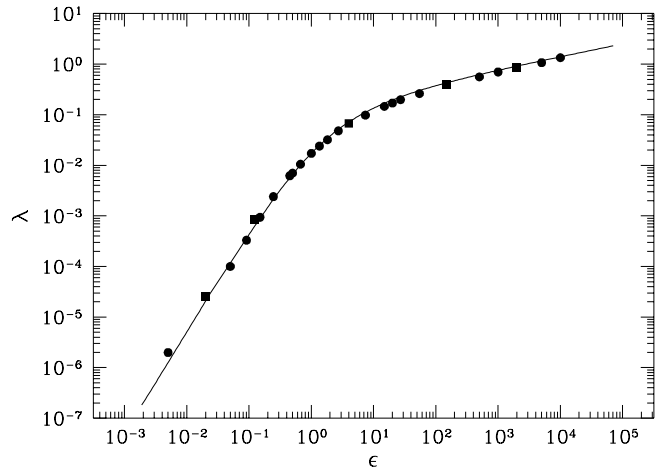,height=8cm,clip=true}}
\caption{FPU-$\beta$ model. Lyapunov exponent $\lambda$ {\it vs.} energy
density $\varepsilon$ with $\beta = 0.1$. The continuous line is
the theoretical computation according to Eq.\
(\protect\ref{lambda_gauss}), while the circles and squares are
the results of numerical simulations with $N$ respectively equal
to 256 and 2000. From Ref. \protect\cite{CCP}.
\label{fig_lyap_fpu} }
\end{figure}

A completely analytical computation of $\lambda(\varepsilon )$ has
been performed -- in the thermodynamic limit -- for the FPU-$\beta$
model (such a result first appeared in \cite{prl95}, then
it was refined in \cite{CCP,physrep}) and for other models. We
report in Fig.\ \ref{fig_lyap_fpu} the result for the FPU case:
the agreement is strikingly good. The analytic values of $\lambda$
agree with the numerical ones with errors of a few percent in a
range of six orders of magnitude both in $\varepsilon =E/N$ and
$\lambda$, and no use of adjustable parameters has been made. A
preliminary explanation of the existence of the SST proceeds as
follows. At low $\varepsilon$, the amplitude of the curvature
fluctuations $\sigma_\Omega$ is much smaller than the average
curvature $\Omega_0$, thus the mechanical manifolds are not very
different from constant curvature manifolds, so that the geodesic
flow has many of the features that it would have if it lived on a
strictly constant curvature (equal to the average curvature)
manifold, and, loosely speaking, a slow phase space filling
through tortuous paths will take place: chaos is weak. Conversely,
when $\sigma_\Omega\sim\Omega_0$, we can imagine that no
similarity at all will exist between the chaotic geodesic flow and
its integrable counterpart living on a constant curvature (equal
to the average curvature) manifold. As a consequence the geodesic
flow can quickly diffuse in any direction in phase space: chaos is
strong.


Other systems for which good results have been obtained are: a
one-dimensional chain of coupled rotators \cite{CCP}, two and
three dimensional classical XY Heisenberg models \cite{CCCP}, two
and three dimensional classical lattice $\varphi^4$ models
\cite{CCP1,CCCPPG}, ``mean-field'' XY model \cite{Firpo}, though
some adjustments are necessary in these cases.

An important remark is in order. The geometrical theory of chaos
aims at explaining what is the origin of chaos in Hamiltonian
systems, and not at providing a recipe for the computation of
Lyapunov exponents. The impressive success of the theory in
analytically computing Lyapunov exponents for the FPU model, means
that we have actually found the right conceptual framework and the
right explanation for the existence of Hamiltonian chaos and
warrants that any effort to further develop the theory is
worthwhile.

One has to keep in mind that the above given analytic formula for
the largest Lyapunov exponent has a limited validity domain: that
of the fundamental assumption of quasi-isotropy of the mechanical
manifolds.
The next step will be that of relaxing the assumption of
quasi-isotropy by letting in nontrivial topology of configuration
space.

\section{Hamiltonian dynamics, phase transitions and topology}

Though the content of this Section could appear to be somewhat far
from the initial FPU problem, we have nonetheless sketched it in
order to remark how fertile, inspiring and far reaching a
systematic investigation of the (once) surprising behavior of the
FPU dynamics has been.

The macroscopic properties of large-$N$ Hamiltonian systems can be
understood by means of the traditional methods of statistical
mechanics. The origin of the possibility of describing Hamiltonian
systems via equilibrium statistical mechanics are the chaotic
properties underlying the dynamics.

Above, we have observed that the crossover in the
$\varepsilon$-dependence of $\lambda$ phenomenologically
corresponds to a transition between weak and strong chaos (SST),
or slow and fast mixing respectively. Thus we have surmised that
this transition has to be the consequence of some ``structural''
change occurring in configuration space, and thus also in phase
space. This dynamical (mild) transition has been observed, we said
above, in many other systems besides FPU. Then, some natural
questions arise: could some kind of dynamical transition between
weak and strong chaos (possibly sharper than the SST found in FPU
models) be the microscopic counterpart of a thermodynamic phase
transition? and if this was the case, what kind of difference in
the $\lambda (\varepsilon)$ pattern would discriminate between the
presence or absence of a phase transition? and could we make a
more precise statement about the kind of ``structural'' change to
occur in configuration space when the SST corresponds to a phase
transition and when it does not?

During the last years, after an earlier attempt in Ref.\
\cite{Butera} where the classical XY model in two dimensions was
considered, and its largest Lyapunov exponent was found to display
some indication of the transition temperature of the
Kosterlitz-Thouless phase transition, there has been a renewed
interest in the study of the behaviour of Lyapunov exponents in
systems undergoing phase transitions, and a number of papers have
appeared: see \cite{CCCP,CCCPPG,CCP1} and
\cite{Duke_pre,DellagoPosch_hd,DellagoPosch_lj,DellagoPosch_XY,
DellagoPosch_hs,Mehra,Ruffo_prl,Firpo,Ruffo_talk,Antoni_rev}.

Two systems have received considerable attention in this
framework: the so-called lattice $\varphi^4$ model, and the
mean-field XY model. The lattice $\varphi^4$ model is described by
the Hamiltonian
\begin{equation}
H =\sum_i \frac{1}{2} p_i^2 + \frac{J}{2} \sum_{\langle i , j
\rangle} (\varphi_{{i}}-\varphi_{j})^2 + \sum_i \left[
-\frac{m^2}{2} \varphi_{i}^2 +\frac{u}{4!}\varphi_{i}^4\right],
\label{hfi4d}
\end{equation}
where the $p_i$ are momenta conjugated to the $\varphi_i$, real
valued scalar variables defined on the sites of a $d$-dimensional
lattice; $m^2$ and $u$ are positive parameters, and the brackets
$\langle i,j\rangle$ stand for nearest-neighbors. This model has a
phase transition at a finite temperature provided that $d>1$.

The mean-field XY model \cite{Antoni} describes a system of $N$
equally coupled planar classical rotators. It is defined by the
Hamiltonian
\begin{equation}
H= \sum_i \frac{1}{2} p_i^2 +\frac{J}{2N}\sum_{i,j=1}^N \left[ 1 -
\cos(\varphi_i - \varphi_j)\right]-h\sum_{i=1}^N \cos\varphi_i\ .
\label{xymf}
\end{equation}
Here $\varphi_i \in [0,2\pi]$ is the rotation angle of the $i$-th
rotator. Defining at each site $i$ a classical spin vector ${\bf
s}_i = (\cos\varphi_i,\sin\varphi_i)$ the model describes a planar
(XY) Heisenberg system with interactions of equal strength among
all the spins. The equilibrium statistical mechanics of this
system is exactly described, in the thermodynamic limit, by
mean-field theory \cite{Antoni}. In the limit $h\to 0$, this
system has a continuous phase transition.

\begin{figure}[ht]
\centerline{\psfig{file=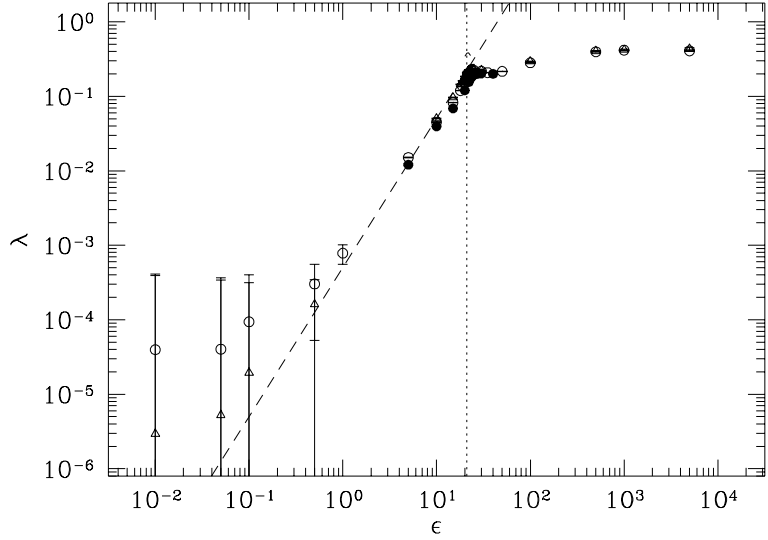,height=8cm,clip=true}}
\caption{Lyapunov exponent $\lambda$ {\em vs.} the energy per
particle $\varepsilon$, numerically computed for the
two-dimensional $O(1)$ $\varphi^4$ model, with $N = 100$ (solid
circles), $N=400$ (open circles), $N=900$ (solid triangles), and
$N=2500$ (open triangles). The critical energy is
marked by a vertical dotted line; the dashed line is the power
law $\varepsilon^2$. From Ref.\ \protect\cite{CCP1}.
\label{fig_lyap_phi4_2d} }
\end{figure}

\begin{figure}[ht]
\centerline{\psfig{file=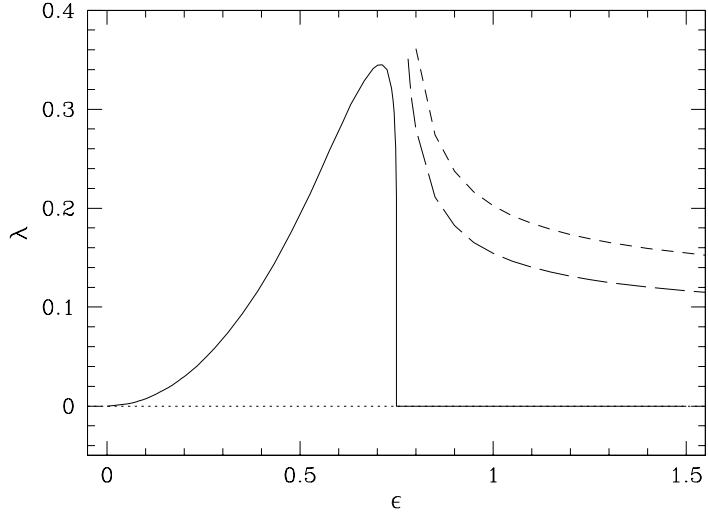,height=8cm,clip=true}}
\caption{Mean field XY model: analytic expression for the Lyapunov
exponent (solid curve). The curves above the transition are
finite-$N$ results for $N=80$ (upper dashed line) and $N=200$
(lower dashed line): here $\lambda \propto
N^{-1/3}$. From Ref.\ \protect\cite{Firpo}. \label{fig_firpo_2} }
\end{figure}

Through standard methods of molecular dynamics, thermodynamical
observables have been computed and found to be in agreement with
statistical mechanical predictions. The energy density
($\varepsilon =E/N$) dependence of the largest Lyapunov exponent
numerically found in the $\varphi^4$ model -- reported in 
Fig.\ \ref{fig_lyap_phi4_2d} -- shows a pattern similar to that found in
the FPU model but now the mild transition between weak and strong
chaos is replaced by an abrupt transition, a sharp SST: a
``cuspy'' point in $\lambda(\varepsilon)$ shows up which
corresponds to the critical energy locating the phase transition.
Also the Lyapunov exponent of the
 mean field XY model,
obtained through an analytic estimate worked out in the limit
$N\rightarrow\infty$ \cite{Firpo} by means of the above discussed
geometrical theory of chaos, sharply signals the phase transition 
(see Fig.\ \ref{fig_firpo_2}).

In both cases, it is evident that the $\varepsilon$-pattern of the
largest Lyapunov exponent clearly signals the presence of a phase
transition; the same happens for all the other models studied in
the above mentioned references.



Then, coming to the other questions, as Lyapunov exponents are
tightly related with the geometry of the mechanical manifolds in
configuration space (as well as in phase space), we have been led
to conjecture that in presence of a phase transition we have to go
to the deeper level of {\it topology} of these manifolds to find
an adequate explanation \cite{FCPS}.
 If this is
actually the case, we are confronted with a possible -- at least
conceptual -- deepening of our understanding of the origin of
phase transitions. In fact, the topological properties of
configuration space submanifolds, mainly equipotential
hypersurfaces $\Sigma_v = V^{-1}(v)=\{ q\in{\mathbb R}^N\vert
V(q)=v\}$ or the regions bounded by them $M_v=\{ q\in{\mathbb
R}^N\vert V(q)\leq v \}$, are already determined when the
microscopic potential $V$ is assigned and are completely
independent of the statistical measures. The appearance of
singularities in the thermodynamic observables could then be the
{\it effect} of a suitable topological transition in configuration
space. Several results strongly support this {\it Topological
Hypothesis} and suggest that a phase transition might well be the
consequence of an abrupt transition between different rates of
change in the topology above and below the critical point. More
details can be found in the review paper \cite{physrep} and in the
subsequent papers: \cite{JSP} where the topology of the $M_v$ is
analytically studied for the mean-field XY model; \cite{EPL,Ptrig}
where the topology of the $M_v$ is analytically studied for a
trigonometric model undergoing also a first-order phase
transition; \cite{PREXY,Ptrig} where an analytic relationship
between topology and thermodynamic entropy is given among other
results; \cite{Thm1,Thm2} where a preliminary account of a general
theorem on topology and phase transitions is given.

\section{Concluding remarks}

By chance, Fermi, Pasta and Ulam chose the initial condition, in their 
numerical experiment on the $\alpha$-model, below the threshold energy of
a transition between regular and chaotic motions. With a stronger initial
excitation, no ``FPU problem'' would have arisen because equipartition of
energy could have been observed even with the rather short integration time
that the authors could afford 50 years ago. Apparently the observed 
phenomenology cannot be explained by the existing formulations of the KAM
theory, both because of the large degree of anharmonicity (nonintegrability)
at which the ST occurs, and because of its slow vanishing at increasing $N$.
Since this threshold energy
goes to zero as the number of degrees of freedom is increased, the FPU
problem is not a true problem for equilibrium statistical mechanics. 
Nevertheless, the existence of possibly long-living transient nonequilibrium 
phenomena draws attention to the relevance of dynamics, initial conditions and
observational time scales in order to assess whether dynamics can be replaced
by statistics or not. 
Because of the cubic potential, the $\alpha$-model is unstable above an upper 
bound in energy density, so FPU considered also the so-called $\beta$-model
which is well defined at any energy. However, in the $\beta$-model it is hard
to detect the ST because it seems to occur at a very low energy density, 
where the convergence of the largest Lypaunov exponent requires huge 
computational times.
On the other hand, the $\beta$-model displays another and much more
interesting chaotic transition, that we called SST, which is a transition
between weak and strong chaos. Strong chaos is related with fast phase 
mixing and fast thermalization of an out of equilibrium initial condition.
Weak chaos is associated with a sudden increase of relaxation times of 
nonequilibrium initial conditions when the energy density is smaller than
a threshold value (which corresponds to the SST). At sufficiently low
energy density the thermalization can be so slow that the system can give
the wrong impression to recur {\it ad infinitum}, if the observational time 
is not long enough.
The study of the $\alpha +\beta$--model, which provides a good approximation
of interatomic interaction potentials of the Morse or Lennard-Jones type,
displays both the ST and the SST. However, only the SST is stable with $N$
and can thus be relevant for equilibrium and nonequilibrium statistical
properties of a large class of classical many-body systems. In fact, this kind
of transition seems a common property of many degrees of freedom Hamiltonian 
systems.

The systematic investigation of the chaotic properties of FPU models -- being
a heavy numerical task -- has
become possible only rather recently, with the advent of modern powerful
computers. The results so obtained demanded  a satisfactory and constructive
explanation of the origin of Hamiltonian chaos as well as for the reason of
the transition between weak and strong chaos.
Motivated by the need of understanding chaos in FPU models, we have started a 
new and successful theory of Hamiltonian chaos which resorts to basic
concepts and methods of Riemannian geometry. Later on, all these findings
have suggested to look at phase transition phenomena from a new point of view
which, eventually, has inspired the development of a new theoretical
approach to them, based on topological concepts.

{\bf Acknowledgments:} We thank D.\ K.\ Campbell, P.\ Rosenau and 
G.\ Zaslavsky for their kind invitation to contribute to this special
issue commemorating the 50th anniversary of the fundamental work
by E.\ Fermi, J.\ Pasta and S.\ Ulam.

E.G.D.C.\ gratefully acknowledges support from the Office of Basic Energy 
Sciences of the US Department of Energy under Grant DE-FG02-88-ER13847.


\end{document}